\documentclass[twocolumn]{aastex631}
\usepackage{natbib}

\def\rpicard{\textsf{\itshape rPICARD}}
\newcommand{\casatask}[1]{\texttt{\textbf{{#1}}}}
\newcommand{\casainput}[1]{\texttt{{#1}}}
\newcommand{\casaparm}[1]{\texttt{\textit{#1}}}

\accepted{by PASP on 27 Sept 2022}

\shorttitle{Common Astronomy Software Applications}
\shortauthors{CASA Team et al.}
\graphicspath{{./}{figures/}}

\begin{document}

\title{CASA, the Common Astronomy Software Applications for Radio Astronomy}

\correspondingauthor{Bjorn Emonts}
\email{casa-feedback@nrao.edu}

\collaboration{48}{The CASA Team}

\author{Ben Bean}
\affil{National Radio Astronomy Observatory, 800 Bradbury Dr., SE Ste 235, Albuquerque, NM 87106, USA}

\author[0000-0001-7164-0089]{Sanjay Bhatnagar}\affil{National Radio Astronomy Observatory, P.O. Box O, Socorro, NM 87801, USA}

\author[0000-0002-4857-3399]{Sandra Castro}\affil{European Southern Observatory, Karl Schwarzschild Strasse 2, D-85748 Garching, Germany}

\author[0000-0002-3106-7676]{Jennifer Donovan Meyer}\affil{National Radio Astronomy Observatory, 520 Edgemont Road, Charlottesville, VA 22903}

\author[0000-0003-2983-815X]{Bjorn Emonts}\affil{National Radio Astronomy Observatory, 520 Edgemont Road, Charlottesville, VA 22903}

\author{Enrique Garcia}\affil{European Southern Observatory, Karl Schwarzschild Strasse 2, D-85748 Garching, Germany}

\author[0000-0002-6116-678X]{Robert Garwood}\affil{National Radio Astronomy Observatory, 520 Edgemont Road, Charlottesville, VA 22903}
	
\author{Kumar Golap}\affil{National Radio Astronomy Observatory, P.O. Box O, Socorro, NM 87801, USA}
	
\author{Justo Gonzalez Villalba}\affil{European Southern Observatory, Karl Schwarzschild Strasse 2, D-85748 Garching, Germany}
	
\author{Pamela Harris}\affil{National Radio Astronomy Observatory, P.O. Box O, Socorro, NM 87801, USA}
	
\author{Yohei Hayashi}\affil{National Astronomical Observatory of Japan, 2-21-1 Osawa, Mitaka, Tokyo 181-8588, Japan}
	
\author{Josh Hoskins}\affil{National Radio Astronomy Observatory, 520 Edgemont Road, Charlottesville, VA 22903}

\author{Mingyu Hsieh}\affil{National Radio Astronomy Observatory, P.O. Box O, Socorro, NM 87801, USA}
	
\author[0000-0002-5825-9635]{Preshanth Jagannathan}\affil{National Radio Astronomy Observatory, P.O. Box O, Socorro, NM 87801, USA}
	
\author[0000-0002-6965-8893]{Wataru Kawasaki}\affil{National Astronomical Observatory of Japan, 2-21-1 Osawa, Mitaka, Tokyo 181-8588, Japan}
	
\author{Aard Keimpema}\affil{Joint Institute for VLBI ERIC, Oude Hoogeveensedijk 4, 7991 PD Dwingeloo, The Netherlands}
	
\author{Mark Kettenis}\affil{Joint Institute for VLBI ERIC, Oude Hoogeveensedijk 4, 7991 PD Dwingeloo, The Netherlands}
	
\author[0000-0002-5648-4206]{Jorge Lopez}\affil{National Radio Astronomy Observatory, 520 Edgemont Road, Charlottesville, VA 22903}

\author{Joshua Marvil}\affil{National Radio Astronomy Observatory, P.O. Box O, Socorro, NM 87801, USA}

\author{Joseph Masters}\affil{National Radio Astronomy Observatory, 520 Edgemont Road, Charlottesville, VA 22903}

\author{Andrew McNichols}\affil{National Radio Astronomy Observatory, 520 Edgemont Road, Charlottesville, VA 22903}
	
\author[0000-0001-7256-9114]{David Mehringer}\affil{National Radio Astronomy Observatory, 520 Edgemont Road, Charlottesville, VA 22903}
	
\author{Renaud Miel}\affil{National Astronomical Observatory of Japan, 2-21-1 Osawa, Mitaka, Tokyo 181-8588, Japan}
	
\author{George Moellenbrock}\affil{National Radio Astronomy Observatory, P.O. Box O, Socorro, NM 87801, USA}
	
\author[0000-0002-2981-5612]{Federico Montesino}\affil{European Southern Observatory, Karl Schwarzschild Strasse 2, D-85748 Garching, Germany}

\author[0000-0003-3780-8890]{Takeshi Nakazato}\affil{National Astronomical Observatory of Japan, 2-21-1 Osawa, Mitaka, Tokyo 181-8588, Japan}
	
\author{Juergen Ott}\affil{National Radio Astronomy Observatory, P.O. Box O, Socorro, NM 87801, USA}
	
\author[0000-0002-8704-7690]{Dirk Petry}\affil{European Southern Observatory, Karl Schwarzschild Strasse 2, D-85748 Garching, Germany}
	
\author{Martin Pokorny}\affil{National Radio Astronomy Observatory, P.O. Box O, Socorro, NM 87801, USA}
	
\author{Ryan Raba}\affil{National Radio Astronomy Observatory, 520 Edgemont Road, Charlottesville, VA 22903}
	
\author{Urvashi Rau}\affil{National Radio Astronomy Observatory, P.O. Box O, Socorro, NM 87801, USA}

\author{Darrell Schiebel}\affil{National Radio Astronomy Observatory, 520 Edgemont Road, Charlottesville, VA 22903}
	
\author{Neal Schweighart}\affil{National Radio Astronomy Observatory, 520 Edgemont Road, Charlottesville, VA 22903}
	
\author{Srikrishna Sekhar}\affil{Inter-University Institute for Data Intensive Astronomy, University of Cape Town, Rondebosch, Cape Town, 7701, South Africa}\affil{National Radio Astronomy Observatory, P.O. Box O, Socorro, NM 87801, USA}
	
\author{Kazuhiko Shimada}\affil{National Astronomical Observatory of Japan, 2-21-1 Osawa, Mitaka, Tokyo 181-8588, Japan}
	
\author{Des Small}\affil{Joint Institute for VLBI ERIC, Oude Hoogeveensedijk 4, 7991 PD Dwingeloo, The Netherlands}
	
\author{Jan-Willem Steeb}\affil{National Radio Astronomy Observatory, 520 Edgemont Road, Charlottesville, VA 22903}

\author{Kanako Sugimoto}\affil{National Astronomical Observatory of Japan, 2-21-1 Osawa, Mitaka, Tokyo 181-8588, Japan}

\author{Ville Suoranta}\affil{National Radio Astronomy Observatory, 520 Edgemont Road, Charlottesville, VA 22903}
	
\author[0000-0002-4298-4461]{Takahiro Tsutsumi}\affil{National Radio Astronomy Observatory, P.O. Box O, Socorro, NM 87801, USA}
	
\author[0000-0001-5473-2950]{Ilse M. van Bemmel}\affil{Joint Institute for VLBI ERIC, Oude Hoogeveensedijk 4, 7991 PD Dwingeloo, The Netherlands}
	
\author[0000-0003-2884-9834]{Marjolein Verkouter}\affil{Joint Institute for VLBI ERIC, Oude Hoogeveensedijk 4, 7991 PD Dwingeloo, The Netherlands}

\author{Akeem Wells}\affil{National Radio Astronomy Observatory, 520 Edgemont Road, Charlottesville, VA 22903}
	
\author{Wei Xiong}\affil{National Radio Astronomy Observatory, 800 Bradbury Dr., SE Ste 235, Albuquerque, NM 87106, USA}

\author[0000-0001-8525-4605]{Arpad Szomoru}\affil{Joint Institute for VLBI ERIC, Oude Hoogeveensedijk 4, 7991 PD Dwingeloo, The Netherlands}

\author{Morgan Griffith}\affil{National Radio Astronomy Observatory, 520 Edgemont Road, Charlottesville, VA 22903}

\author{Brian Glendenning}\affil{National Radio Astronomy Observatory, P.O. Box O, Socorro, NM 87801, USA}

\author{Jeff Kern}\affil{National Radio Astronomy Observatory, 520 Edgemont Road, Charlottesville, VA 22903}



\begin{abstract}

CASA, the {\sl Common Astronomy Software Applications}, is the primary data processing software for the Atacama Large Millimeter/submillimeter Array (ALMA) and the Karl G. Jansky Very Large Array (VLA), and is frequently used also for other radio telescopes. The CASA software can handle data from single-dish, aperture-synthesis, and Very Long Baseline Interferometery (VLBI) telescopes. One of its core functionalities is to support the calibration and imaging pipelines for ALMA, VLA, VLA Sky Survey (VLASS), and the Nobeyama 45m telescope. This paper presents a high-level overview of the basic structure of the CASA software, as well as procedures for calibrating and imaging astronomical radio data in CASA. CASA is being developed by an international consortium of scientists and software engineers based at the National Radio Astronomical Observatory (NRAO), the European Southern Observatory (ESO), the National Astronomical Observatory of Japan (NAOJ), and the Joint Institute for VLBI European Research Infrastructure Consortium (JIV-ERIC), under the guidance of NRAO.

\end{abstract}

\keywords{methods: data analysis -- instrumentation: interferometers -- techniques: image processing -- techniques: imaging spectroscopy -- techniques: interferometric}


\section{Introduction}

Radio astronomy is a discipline that heavily relies on computational resources to image the sky at wavelengths ranging from roughly 10m - 300$\mu$m \citep[e.g.,][]{con16,tho17}. The Common Astronomy Software Applications (CASA)\footnote{\url{https://casa.nrao.edu}} \citep{mcm07} is a software package that enables the calibration, imaging, and analysis of data produced by world-leading radio telescopes.

CASA consists of open-source software for the processing of single-dish and radio interferometric data \citep{jen58}. It consists of a suite of applications implemented in the C++ programming language \citep{str97} and accessible through an Interactive Python interface \citep{per07}. The origin of CASA lies in the AIPS++ project \citep{gle96,mcm06}, which was started in 1992 as the successor of the Astronomical Information Processing System (AIPS) software package \citep{gre03}. The original AIPS++ project was run by a consortium of astronomical institutes, including the Australia Telescope National Facility (ATNF), the National Center for Supercomputing Applications  (NCSA) at the University of Illinois, Jodrell Bank Observatory (JBO), the MERLIN/VLBI National Facility (MERLIN/VLBI), the National Radio Astronomy Observatory (NRAO), and the Netherlands Foundation for Research in Astronomy (ASTRON).\footnote{Initial members also included the National Centre for Radio Astrophysics (NCRA) in India and Canada's Herzberg Institute of Astrophysics (HIA).} In 2004, the AIPS++ code was re-organized and migrated to CASA, and the scripting language was changed from Glish \citep{pax93,schie96} to Python bindings known as `casapy'. At the same time, the core of the AIPS++ libraries formed Casacore \citep{die94,cas19}.\footnote{\url{https://casacore.github.io}} Casacore has offered a stable and nearly static platform for many radio synthesis telescopes. It includes a table based storage mechanism designed for astronomical data, a visibility storage framework, fundamental C++ data structures, an image storage format, numerical methods for astronomy, a library for unit and coordinate conversion, and methods to handle images in the Flexible Image Transport System (FITS) format \citep{pen10} using wcslib for coordinate conversions \citep{cal11}. A consortium led by NRAO started to develop the CASA package, primarily aimed at supporting the Atacama Large Millimeter/submillimeter Array (ALMA) \citep{woo09} and the Karl G. Jansky Very Large Array (VLA) \citep{tho80,cha14}, but on a best-effort basis also other radio telescopes. CASA layers advanced calibration, imaging and image analysis along with basic visualization and telescope-specific support on top of the Casacore base. Currently, CASA is the primary data processing software for ALMA and the VLA, and through the versatility of the software and external development collaborations, it is commonly used also for other aperture-synthesis and single-dish radio telescopes. 

This paper provides a high-level overview of the CASA software. It is not intended as a complete overview of CASA functionality or of radio processing techniques, but instead allows the readers to familiarize themselves with the principles and philosophy behind the CASA software. In Sect.\,\ref{sec:software} we summarize the structure of CASA, the CASA data model, the Application Programming Interface (API), and the comprehensive CASA documentation. Sect.\,\ref{sec:processing} gives an overview of the different stages of data processing in CASA, both for interferometric and single-dish data. One of the core functionalities of CASA is to support the data calibration and imaging pipelines for ALMA, VLA and the VLA Sky Survey (VLASS), and we briefly discuss this in Sect. \ref{sec:pipelines}. The CASA development process is summarized in Sect.\ref{sec:devprocess}. Sect. \ref{sec:ngcasa} mentions the tentative design of a next-generation CASA package, suitable for data processing with next-generation radio telescopes.

\section{The CASA software}
\label{sec:software}

\subsection{CASA structure and Python}
\label{sec:structure}

\begin{figure}
\centering
\includegraphics[width=0.42\textwidth]{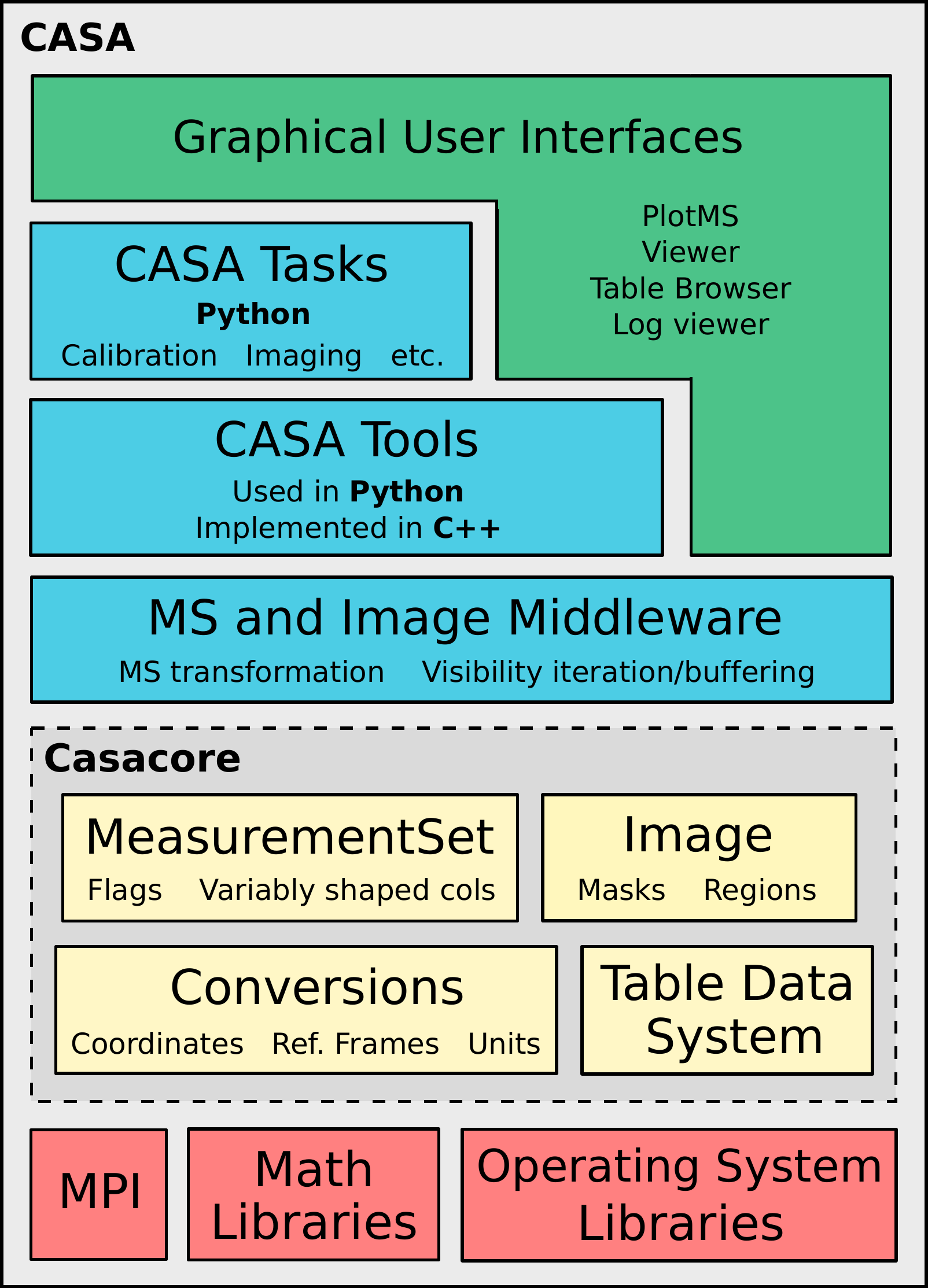}
\caption{Layer architecture diagram of the monolithic CASA software package. The various terminology is discussed throughout Sect.\,\ref{sec:software}}
\label{fig:architecture}
\end{figure}

CASA consists of an open-source code library implemented in C++ \citep{str97}, with some parts in Fortran \citep{bac56}. It is designed to run on Unix platforms, including certain versions of Linux and macOS. Python is used for scripting and to interact with the software through an Interactive Python (IPython) interface \citep{ros95,per07}. Therefore, CASA uses the standard Python syntax to define variables, lists, indices, etc. CASA has traditionally been distributed as a `monolithic', integrated application, including a Python interpreter and all the libraries, packages and modules (Fig.\,\ref{fig:architecture}). From version 6 onward, CASA is also available in `modular' version through pip-wheel installation, which was introduced together with the switch from Python 2 to 3. This provides users with the flexibility to use CASA in their own Python environment.

Initial versions of the CASA top-level user interface were object-oriented, but it was soon adapted to provide a functional, aggregate interface that is easier for users to handle and reminiscent of the AIPS model. The object-oriented interface remained on the layer below. Since then, within CASA, the suite of applications consists of basic objects called \casatask{tools} that can be called to perform operations on the data, as well as user-friendly \casatask{tasks} (see Sect. \ref{sec:tasktools}). CASA tasks are small to medium-size applications that use the CASA tools, and that are built as Python functions with a well-defined set of parameters. In this way, both tools and tasks are Python interfaces to the C++ application layer that implements the more computing-intensive science algorithms and features (Fig.\,\ref{fig:architecture}). Further underneath this application layer is the C++ code framework of Casacore, which is an independent package that contains the core data- and image-handling infrastructure \citep{cas19}. 

The CASA software is designed to facilitate manual interactive and scripted use, as well as batch and pipeline processing. It provides a framework to parallelize processing using multiple cores and computing nodes where available. The history of the present CASA session  is displayed in the CASA \casatask{logger} graphical user interface (GUI), and also stored on disk in a {\sc casa.log} file. 

\subsection{The MeasurementSet and Data Model}

The MeasurementSet (MS) is the database in which interferometric or single-dish data are stored for processing in CASA \citep{wie96,kem00}. The MS is essentially a relational database consisting of a Main table containing the bulk data, and a number of subtables which store meta-data referenced by the Main table to avoid redundancy and to allow database normalization \citep{cod70}. In CASA, the MS is implemented based on the Casacore Table Data Sytem \citep[CTDS;][]{die15,die20}, which is a storage management system that stores each table as a file system directory with a standard set of files, in particular one file per table column. The Main table bulk data can either consist of the interferometric visibilities or single-dish total-power measurements. The Main table rows contain these data chronologically for each time-step, phase-reference direction, spectral window, and baseline, with each row containing a 2-dimensional array that represents the polarization and spectral channel axes. The meta-data consist of entries that identify and characterize the recorded data, such as the spectral window parameters or the positions of the antennas \citep[see][]{kem00}.

The MS Main table initially only contains one column with the radio data, named \texttt{DATA} for interferometric data or \texttt{FLOAT\char`_DATA} for single-dish data. When a calibration is applied to the \texttt{DATA} column, a \texttt{CORRECTED\char`_DATA} column is created to contain the calibrated data, leaving the original data untouched. Other columns can be added, such as the \texttt{MODEL\char`_DATA} column, which stores the expected visibility values that represent an image-plane source model. Other standard columns worth mentioning are the \texttt{FLAG} column, which indicates the data flags (Sect.\,\ref{sec:flagging}); the \texttt{SIGMA} column, which gives the per-channel noise; and the \texttt{WEIGHT} column, which contains the data weights to be used, e.g., when combining different visibilities in imaging.

\subsection{Application Programming Interface (API)}
\label{sec:API}

CASA provides a Python framework, which includes Python packages that expose CASA's functionality in a functional manner (\casatask{casatasks}) and in an object oriented manner (\casatask{casatools}). Additional packages provide basic visualization tools. As mentioned in Sect.\,\ref{sec:structure}, these packages can be installed directly as part of the user's normal Python environment, or the user can download a monolithic version of CASA (Fig.\,\ref{fig:architecture}), which includes a full Python distribution including an extended IPython shell with task invocation  (\casainput{inp}/\casainput{go}) and other extensions. These elements constitute the Application Programming Interface (API) of CASA, which gives users the possibility to configure and extend CASA. An extensive suite of automated tests is used to verify this API for each CASA release.

\subsubsection{CASA configuration and shell}

CASA accepts a variety of configuration options, specified either through configuration files or command line arguments. A configuration file {\sc config.py} can be edited prior to starting a CASA session, which can be used to set reference data paths, pipeline installation, options for display of GUIs, and more. In addition, a {\sc startup.py} file can be used to customize the CASA shell, which provides the environment for interactive Python-based data analysis using CASA tasks and tools, as described in Sect. \ref{sec:tasktools}. The {\sc startup.py} file allows setting paths, importing Python packages, and executing other Python code to prepare the session.

\subsubsection{CASA tasks, tools, and GUIs}
\label{sec:tasktools}

CASA tools are Python objects that use C++ functions to perform operations on the data. The collection of tools contains a large part of the functionality of CASA, but each tool in itself performs a separate low-level operation. This means that procedures to achieve a typical data analysis objective, such as performing a bandpass calibration, typically need many calls to different tools. To make CASA more user-friendly, higher-level CASA tasks are offered, which are constructed based on the tools and cover a comprehensive set of use-cases for data processing and analysis. A CASA task performs a well defined step in the processing of the data, such as loading, plotting, flagging, calibrating, imaging, or analyzing the data. Each task has a well defined purpose and set of input parameters, which may have a two-layer hierarchy containing sub-parameters. 

While CASA tools are Python class objects with methods, CASA tasks can be called as a function with one or more arguments specified, which allows for robust scripting and pipeline use in Python. Where possible and useful, parameters have a specified default value. CASA tasks can also be controlled by setting global parameters. These can be inspected in the terminal via the \casainput{inp} command, before executing the task by typing \casainput{go}. Throughout this paper, we will refer to the names of CASA tasks and tools in \casatask{bold}, and task parameters in \casaparm{italic}. For an overview of the tasks and tools that are available in CASA, see CASA Docs.\footnote{\url{https://casadocs.readthedocs.io}}

CASA also contains a variety of applications with Graphical User Interfaces (GUIs) to examine visibility data, image products, and meta-data. Two of the most widely used GUIs are \casatask{plotms}, for diagnostic plotting of visibility and calibration-table data, and the CASA \casatask{viewer}, for visualizing image products. More information on data analysis and the use of GUIs in CASA is provided in Sect.\,\ref{sec:analysis}.

\subsubsection{External data repository}

Each CASA version comes with a minimal repository of binary data that is required for CASA to function properly. In particular the CASA Measures system, which handles physical quantities with a reference frame and performs reference frame conversions, requires valid Earth Orientation Parameters (EOPs) and solar system object ephemerides. These are contained in the `Measures tables'. The EOP tables are updated daily by ASTRON,\footnote{\url{https://www.astron.nl}} based on the geodetic information from services like the International Earth Rotation and Reference Systems Service (IERS).\footnote{\url{https://www.iers.org}} The ephemeris tables are copied over from the {\it Horizons} online solar system data and ephemeris computation service at the Jet Propulsion Laboratory (JPL).\footnote{\url{https://ssd.jpl.nasa.gov/horizons}} Other data that is updated less frequently is also stored in the CASA data repository, such as observatory-specific beam models, correction tables for Jy/K conversion, and files that specify the antenna array configurations for the CASA simulator.

The CASA data repository, and the runtime data therein, is necessarily updated frequently compared to the much sparser schedule by which new CASA versions are released. Operations on recently observed data may need an up-to-date data repository to work correctly. 
Upon starting a CASA session, it may therefore be necessary to update either the Measures tables or the entire data repository with the most recent version. Routines for handling these external data dependencies are readily available in CASA.

\subsection{Parallel processing}
\label{sec:parallel}

In order to achieve high processing speeds, CASA provides a framework to run tasks and commands in parallel using multiple cores and computing nodes \citep{rob99,cas17}. The CASA Parallelization framework is implemented using the Message Passing Interface (MPI) standard \citep{mpi93}. The CASA distribution comes with a wrapper of the MPI executor called \casainput{mpicasa}, which configures the environment to run CASA in parallel. CASA implements parallelization both at the Python and C++ level. Calibration and manipulation tasks, as well as continuum imaging, use Python-level parallelization, while spectral cube imaging uses specific C++ level parallelization.

Imaging in CASA offers parallelization by partitioning along the time axis (for continuum imaging) or frequency axes (for spectral cube imaging), which shortens the overall execution time of gridding, deconvolution, and auto-masking. Certain parts of imaging also use multi-threading to improve performance, e.g. in Fourier transforms. 

Several calibration and data-manipulation tasks can be executed in a trivially parallel mode where there is little or no dependency or need for communication between the parallel processes. So far, parallel execution of one CASA session in multiple processes has to be specifically initiated by starting the session using \casainput{mpicasa} instead of \casainput{casa}. However, it is foreseen that in the future parallel processing will become the standard way of reducing data (see Sect. \ref{sec:ngcasa}).

\subsection{CASA Documentation}
\label{sec:casadocs}

The official CASA documentation is the online library of CASA Docs.\footnote{\url{https://casadocs.readthedocs.io}} A version of CASA Docs is published with each official CASA release. CASA Docs describes in detail the functionality of the CASA software. For each task, CASA Docs provides an extensive description, as well as an overview of the parameters that are available for that specific task. The CASA Docs page with the task description and parameter overview can also be opened from within CASA with the command \casainput{doc(`taskname')}. CASA Docs further provides detailed background information, installation instructions, useful examples, release information, and a list of known issues. CASA Docs also includes a CASA Memo and Knowledgebase section.

We note that CASA Docs is not intended to teach users all the fundamentals of radio astronomy. For that, a number of excellent books and review articles are available \citep[e.g.,][]{chr85,tho86,cla95,tay99,wil13,con16}. There is also a comprehensive collection of online tutorials, or `CASA Guides', which provide step-by-step recipes for processing various types of data sets in CASA.\footnote{\url{https://casaguides.nrao.edu}} Various telescope teams maintain these data processing tutorials for ALMA, VLA, Australia Telescope Compart Array (ATCA; \citealt{fra92}), European VLBI Network (EVN; \citealt{zen15}), and CASA simulations. The CASA website\footnote{\url{https://casa.nrao.edu}} contains links to all the information and documentation that CASA offers, including how to download CASA.

\section{Data processing}
\label{sec:processing}

\begin{figure}
\centering
\includegraphics[width=0.47\textwidth]{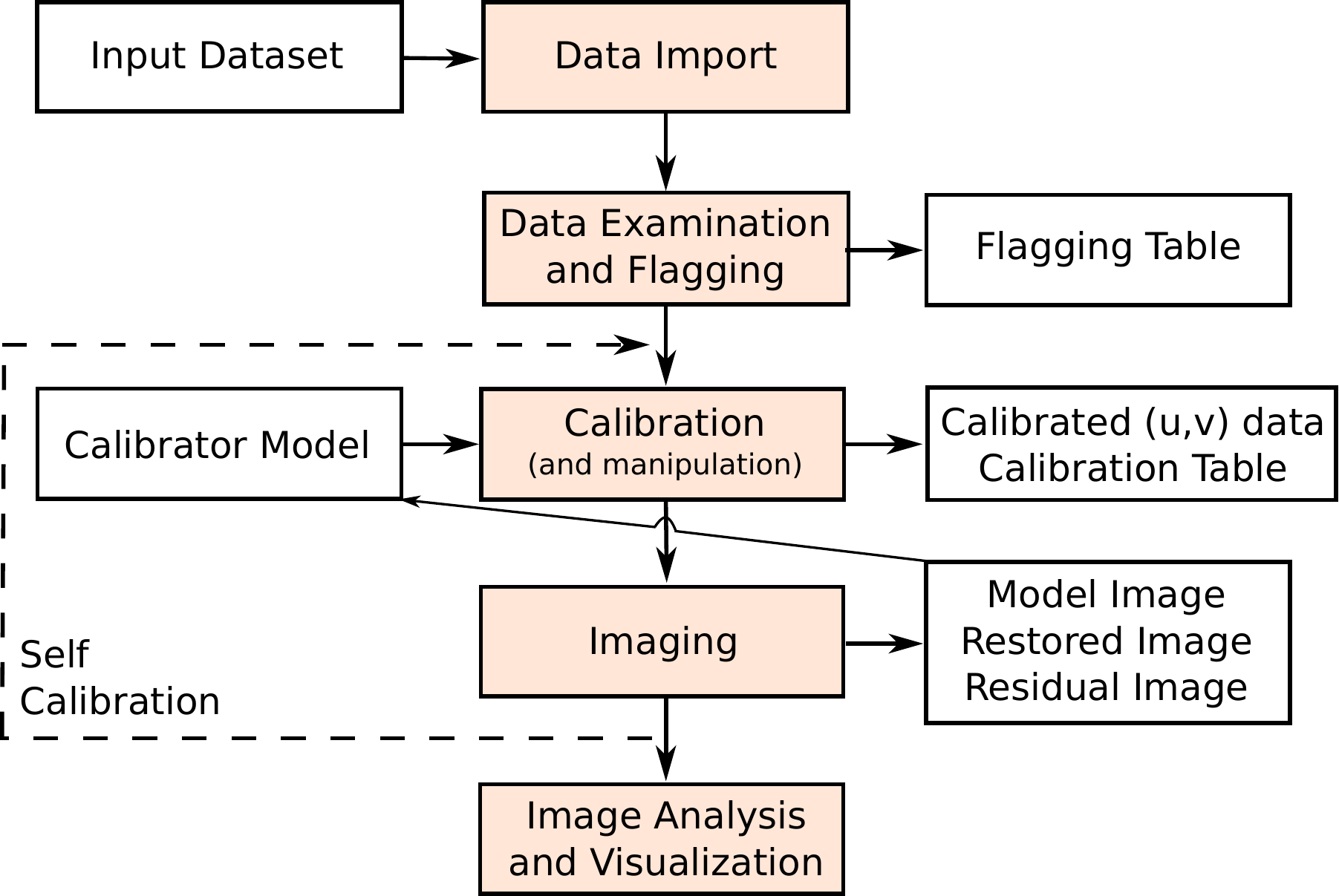}
\caption{Flow chart of the basic operations that a general user will carry out in a typical end-to-end CASA data processing session.}
\label{fig:flowchart}
\end{figure}

Figure \ref{fig:flowchart} shows a typical end-to-end workflow diagram for CASA data processing for radio interferometric data. In this Section, we will discuss the different stages of the data processing, including data import and export (Sect. \ref{sec:import}), data examination and flagging (Sect. \ref{sec:flagging}), calibration (Sect. \ref{sec:calibration}), data manipulation (Sect. \ref{sec:manipulation}), imaging (Sect. \ref{sec:imaging}), visualization and data analysis (Sect. \ref{sec:analysis}), and simulations (Sect. \ref{sec:simulations}). We will address how these stages in the data processing are handled by CASA.

\subsection{Data processing I: Import and export}
\label{sec:import}

For importing and exporting data, CASA can handle various file formats. The raw data from ALMA and the VLA are not recorded in an MS with CASA tables, but instead in the Astronomy Science Data Model (ASDM) format, also referred to as the ALMA  Science Data Model or Science Data Model (SDM) \citep{vial06}. The ASDM is an extension of the MS, but also contains all the observatory-specific meta-data. It is implemented in a hybrid format using `Binary Large Objects' for the bulk data and large meta-data subtables, and XML for smaller meta-data subtables. When processing ALMA or VLA data, the first step has to be the import of the (A)SDM into the CASA MS format using the task \casatask{importasdm}. Data from a few other telescopes and astronomical software packages can also be directly imported into a CASA MS, as can any general radio astronomy data stored in the FITS formats UVFITS and FITS-IDI \citep{wel81,pen10,gre16,gre20}.

\subsection{Data processing II: Examination and Flagging}
\label{sec:flagging}

To examine data, CASA offers a range of tasks and GUIs. Task \casatask{listobs} provides a summary of the contents of an MS as ASCII text. A variety of GUI-based tasks and applications permit to plot and select visibilities and meta-data, often interactively or 
through a scriptable interface. These include \casatask{plotants}, which plots the antenna positions that were used during the observations; \casatask{browsetable}, which permits to inspect and edit the contents of any CASA table and its meta-data; and \casatask{plotms}, which interactively creates diagnostic plots from MSs and calibration tables. The \casatask{ms} and \casatask{tb} tool interfaces allow the extraction of data chunks into  numpy arrays, for further examination and visualization using standard Python libraries.

An important application for the visualization and selection of $(u,v)$-data is `flagging'. The term `flagging' or `applying flags' means discarding data that should not
be included in the calibration or subsequent scientific analysis, because the data is compromised by
known systematic errors or technical issues that happened during observations. These could for example be data from an antenna which line-of-sight to the source was partially blocked (`shadowed') by another antenna, or data from spectral channels at the edge of the receiver band that are particularly noisy. Another important reason for flagging of $(u,v)$-data is radio frequency interference (RFI; \citealt{fri01}, \citealt{ell05}, \citealt{bri05}, see also \citealt{tho17}, and references therein). RFI consists of strong, predominantly human-made radio emission from electrical equipment, such as satellites, television towers, cars, or mobile devices, which contaminates the faint radio signals from astronomical objects. RFI is often limited to well-defined ranges in frequency or time, and the corresponding visibilities can be discarded by flagging the data \citep[e.g.,][]{mid06,off10}. 

CASA offers several ways to apply flags. Besides the GUI-based applications described above, which permit easy graphical selection and flagging of a sub-set of the visibility data, there is also the all-purpose flagging task \casatask{flagdata}. The task \casatask{flagdata} relies on command-line input for manual data selection and automated algorithms to apply flags. For manual data selection, users can specify antennas, baselines, spectral windows, shadowing limits, and other parameters. In addition, \casatask{flagdata} includes the following automated flagging algorithms:
\begin{itemize}
\item{\casaparm{clip:} flags visibilities associated with ranges in data values. This includes values in the meta-data, such as water vapor radiometer (WVR) or system temperature (T$_{\rm sys}$) measurements.}
\item{\casaparm{tfcrop:} creates time-averaged spectra of the visibility amplitudes for each field, spectral window, timerange, and baseline. It then fits a polynomial to the bandshape of each spectrum, and subsequently identifies and flags data points that deviate from this polynomial fit. \casaparm{tfcrop} is optimized for flagging strong, narrow-band RFI}.
\item{\casaparm{rflag:} detects outliers based on sliding-window root-mean-square (rms) filters. This means that as the algorithm iterates through the data in chunks of time, it calculates statistics across the time-chunks and applies flags based on user-supplied thresholds.} 
\end{itemize}
Flagging does not actually delete data in the MS, but just makes entries in corresponding Boolean arrays inside the MS. A version of this \texttt{FLAG} column can be copied over to a CASA flag versions table, under a separate directory with the extension {\sc <msname>.flagversions}. This provides backups of the flags, which can be restored to the MS from which they were created, in order to get back to a previous flag version. The CASA task \casatask{flagmanager} allows users to manage different versions of flags in the data.

\subsection{Data processing III: Calibration}
\label{sec:calibration}

\subsubsection{Interferometric calibration}

Calibration is the process of correcting the signals measured with an interferometer or single-dish radio telescope for instrumental and environmental propagation factors that corrupted the desired astronomical signal, and transform the data from instrumental units to absolute standard units. This is necessary for making an accurate image of the sky, which can be related to images made by other telescopes and theoretical predictions. In CASA, the calibration process consists of determining a series of complex correction factors and applying these corrections to the visibility measurements of objects of scientific interest. Calibration solutions are typically derived from observations of well-characterized calibrator sources, often radio-bright and unresolved quasars for which a simple visibility model may be assumed, and occasionally solar-system objects for absolute flux calibration. In cases where a source in the target-field is bright enough, calibration solutions can often also be obtained or improved through the iterative process of self-calibration, using this target source as the de facto calibrator source \citep[e.g.,][]{rea84,wil89,cor99}. 

The MeasurementSet structure in CASA is designed to permit interferometric data to be calibrated following the Measurement Equation \citep{noo96,ham96,sau96_1}. The Measurement Equation is based on the fact that the visibilities measured by an interferometer have been corrupted by a sequence of multiplicative factors, arising from the atmosphere, antennas, electronics, correlator, and downstream signal-processing. For visibility calibration in CASA, the general Measurement Equation \citep{ham96,sau96_1} can be written as:
\begin{equation}
\vec{V}_{ij}\,=\,J_{ij}\,\vec{V}^{\,\text{TRUE}}_{ij},
\label{eqn:measurementequation}
\end{equation}
where $\vec{V}_{ij}$ represents the observed visibility, which is a vector of complex numbers that represents the amplitude and phase of the raw correlations formed among the sampled dual-polarization wavefronts, as received by a pair of antennas ($i$ and $j$) at each time sample and per spectral channel.  $\vec{V}^{\text{{TRUE}}}_{ij}$ represents the corresponding true visibilities, in a nominally perfect polarization basis (typically either linear or circular), that are to be recovered by the calibration process, and which are proportional to combinations of the true Stokes parameters of the source visibility function \citep{sto52}.  $J_{ij}$ is an operator that represents an accumulation of all corruption factors affecting the correlations on baseline $i-j$.  As written here, $J_{ij}$ is a Mueller matrix for baseline $i-j$, formed in most cases from the outer product of a pair of (single-index) antenna-based Jones matrices, $J_{i}$:
\begin{equation}
    J_{ij}\,=\,J_{i}\otimes\,J_{j}.
\end{equation}
Each antenna-based Jones matrix, $J_{i}$, models the propagation of a signal from a radiation source through to the voltage output of antenna $i$, and characterizes the net effect on the resulting correlations \citep{jon41,mue48,hei01}. Factoring $J_{ij}$ in equation \ref{eqn:measurementequation} into the most common series of recognized effects, we have: 
\begin{equation}
\vec{V}_{ij}\,=\,M_{ij}\,K_{ij}\,B_{ij}\,G_{ij}\,D_{ij}\,E_{ij}\,P_{ij}\,T_{ij}\,F_{ij}\,\vec{V}^{\,\text{TRUE}}_{ij},
\label{eqn:measurementequation2}
\end{equation}
\noindent with the individual terms described in detail below. For just the antenna-based calibration terms, we can write:
\begin{equation}
J_{i}\,=\,K_{i}\,B_{i}\,G_{i}\,D_{i}\,E_{i}\,P_{i}\,T_{i}\,F_{i}.
\label{eqn:jonesfactors}
\end{equation}

\noindent The order of terms in equation \ref{eqn:jonesfactors} reflects the order in which the incoming signal encounters the corrupting effects (from right to left), and in general, terms cannot be arbitrarily reordered, i.e., they do not all algebraically commute.  In practice, not all terms need to be considered in all cases, and some (or even most) may be ignored depending upon their relative importance with respect to dynamic range requirements and scientific goals. In aggregate, the calibration terms in the Measurement Equation describe a net effective (imperfect) polarization basis, characterizing the departure from the ideal intended polarization basis of the instrument. The internal algebra and properties of particular calibration terms in some cases depend on the nominal basis, as will the heuristics engaged to solve for or calculate them, and thus also their implementations in CASA. However, the form of the equation as expressed here represents an approximation to the sequence of physical effects encountered by the observed wavefront, and is therefore polarization basis-agnostic.

Solving for an antenna-based factor is generally an over-determined problem, since there are only $N_{\text{ant}}$ antenna-based factors for $N_{\text{ant}}(N_{\text{ant}}-1)/2$ baselines, per term.  The specific calibration terms recognized within the CASA calibration model are as follows:
\begin{itemize}
\item{$F_{i}$: Ionospheric effects, including dispersive delay and Faraday rotation \citep[e.g.,][]{gas18}.  Most relevant at low radio frequencies ($\la$5 GHz), and typically estimated from information on the ionospheric total electron content.}
\item{$T_{i}$: Tropospheric effects, such as opacity and path-length variations \citep[e.g.,][]{hin71}, which stochastically affect visibility amplitude and phase, respectively, in a polarization-independent manner. Relevant at all radio frequencies, but most important at higher frequencies, where tropospheric variations set the timescale for calibration. Typically solved from the visibility data themselves, on a calibrator observed periodically near the science target. Estimates may also be obtained from water vapor radiometry (WVR; \citealt{roc91,emr09,mau17}).}
\item{$P_{i}$: Parallactic angle rotation, which describes the orientation of the antennas' polarization in the coordinate system of the sky.  Calculated analytically from observational geometry information.}
\item{$E_{i}$: Effects introduced by properties of the optical components of the telescopes, including gain response as a function of elevation and the net aperture-efficiency scale.}
\item{$D_{i}$: Instrumental polarization response, describing the polarization leakage between feeds. This factor describes the fraction of one hand of nominally received polarization that is detected by the receptor for the other hand, and vice versa.  Solved from observations on an appropriately chosen calibrator. }
\item{$G_{i}$: General time- and polarization-dependent complex gain response, including phase and amplitude variations due to the signal path between the feed and the correlator, and sometimes including tropospheric and ionospheric effects when not separately factored into $F_{i}$ and $T_{i}$.  Solved from the visibility data themselves (like $T_{i}$), $G_{i}$ includes the scale factor for absolute flux density calibration, by referring solutions to observations of a calibrator with known flux density. In some cases, this scale reconciliation is performed on $T_{i}$ solutions.}
\item{$B_{i}$: The amplitude and phase `bandpass' response introduced by spectral filters and other components in the electronic transmission of the signal. This is effectively a frequency channel-dependent version of $G_{i}$.  Usually assumed to be stable in time, $B_{i}$ is solved from calibrator observations with sufficient signal-to-noise ratio per frequency channel.}
\item{$K_{i}$: General geometrically-parametrized gain phases that are time- and frequency-dependent, such as delay and delay-rate.  Includes antenna-position corrections and traditional fringe-fitting.}
\item{$M_{ij}$: Baseline-based correlator (non-closing) errors.  When used with extreme care, baseline-based solutions can account for residuals not factorable as antenna-based errors, but this usually indicates a failure to fully and adequately model and calibrate more subtle antenna-based effects, such as instrumental polarization.  Once invoked, subsequent antenna-based calibration cannot be further used reliably since the baseline-based term will have absorbed antenna-based information arbitrarily.}
\end{itemize}

Occasionally, specializations of some of these terms are invoked for particular purposes. For example, system temperature calibration is implemented as a variation of bandpass calibration, position angle calibration is implemented as an offset to the $P_{i}$ term, and cross-hand phase as an offset to the $G_{i}$ term (but located in front of $D_{i}$).  In general, the modular implementation of calibration terms in CASA is designed to permit such specializations, where warranted.  Additional corrections, such as for direction-dependent widefield or wideband effects, can optionally be invoked during the imaging stage (see Sect. \ref{sec:gridding}).\footnote{Treatment of direction-dependent effects within the Measurement Equation for radio interferometers is discussed by \citet{smi11a,smi11b} and \citet{tas14}.}

\subsubsection{Calibration methodology in CASA}
\label{sec:calmethod}

In CASA, solutions for the calibration corrections can be derived using various calibration tasks and tools. The calibration solutions are then stored in separate calibration tables, in a system similar to the MeasurementSet. The most widely used tasks for deriving calibration corrections are:
\begin{itemize}
\item{\casatask{gaincal}: solves for time- and optionally polarization-dependent (but channel-independent) variations in the complex gains (`phases' and/or `amplitudes'), i.e., $G_{i}$ and $T_{i}$.   Also used to derive rudimentary delay corrections, as a variation of $K_{i}$.}
\item{\casatask{bandpass}: solves for frequency-dependent complex gains, i.e., $B_{i}$.}
\item{\casatask{fluxscale}: applies a scale factor to the $G_{i}$ or $T_{i}$ gain solutions from \casatask{gaincal}, according to the gains derived from observations of the flux-density calibrator, on the assumption that the net electronic gain is stable with time. Normally the task \casatask{setjy} is used to set the model visibility amplitude and phase associated with a flux density scale prior to running \casatask{gaincal} and \casatask{fluxscale}.}
\item{\casatask{polcal}: solves for instrumental polarization calibration factors, including leakage, cross-hand phase, and position angle corrections.}
\item{\casatask{fringefit}: solves for delay- and rate-parameterized gain phases for cases where uncertain array geometry and/or distinct clocks limit coherence in frequency and time, as is common in Very Long Baseline Interferometry (VLBI; e.g., \citealt{wie11}).  Also supports a dispersive delay term.  The implementation of \casatask{fringefit} is described in detail by \cite{bem22}.}
\item{\casatask{gencal}: derives various calibration solutions from ancillary information stored in the MeasurementSet or otherwise specified or retrieved, including ionosphere, system temperature, antenna position corrections, opacity, gain curves, etc.}
\item{\casatask{applycal}: applies all specified calibration to the MeasurementSet \texttt{DATA} column, according to the order prescribed in the Measurement Equation (Eqn. \ref{eqn:measurementequation2}), and writes out the \texttt{CORRECTED\char`_DATA} column for imaging.} 
\end{itemize}
A range of other tasks and tools for calibration support are available in CASA, including \casatask{plotweather} to plot weather information and estimate opacities, \casatask{plotbandpass} for plotting bandpass information, \casatask{wvrgcal} for WVR-derived gains \citep{nik12}, and \casatask{blcal} to derive baseline-based gains.

Solving for calibration is a generalized bootstrapping process wherein each additional solved-for component is determined relative to the best available existing information for other calibration terms and the calibrator visibility model.  Initially, there may be no prior calibration available (or only a few terms derived from ancillary information), and only point-like visibility models.  As new calibration is derived, it may be desirable to revise terms that were solved for earlier in the process. For example, a solution for the bandpass, $B_{i}$, may be determined with a provisional time-dependent $G_{i}$ solution from the same calibrator data (but derived from only a subset of frequency channels without the benefits of a prior $B_{i}$ solution), and then the $G_{i}$ solution will be revised and improved using the $B_{i}$ solution as a prior and all frequency channels for more sensitivity.  Algebraically, the prescribed order of calibration terms (\ref{eqn:jonesfactors}) is respected in all solves.  When all relevant calibration terms have been solved-for, their aggregate is applied (with appropriate interpolation) to calibrator and science target data for imaging and deconvolution.  

Insofar as the calibration derived from calibrator observations may not precisely characterize corruptions occurring in the science target data (due to time- and direction-dependence), it is often desirable to {\em self-calibrate} against the visibility model derived from the initial imaging and deconvolution process, and then re-image to better bootstrap and converge on an optimized net calibration and source model \citep[e.g.,][]{rea84,wil89,cor99}. In some cases, the calibrators may not be perfect point sources, and self-calibration may be used to jointly optimize their visibility model(s), and the calibration derived from them, before proceeding to the science target.  The CASA implementation for visibility calibration is designed to support this generalized self-calibration ideal for the iterated revision and convergence of all calibration terms and source model estimates.  In practice, the largest benefit is obtained by iterative revision of time-dependent gains, mainly due to the troposphere, and the visibility model of the source.

\begin{figure*}[tb!]
\centering
\includegraphics[width=0.85\textwidth]{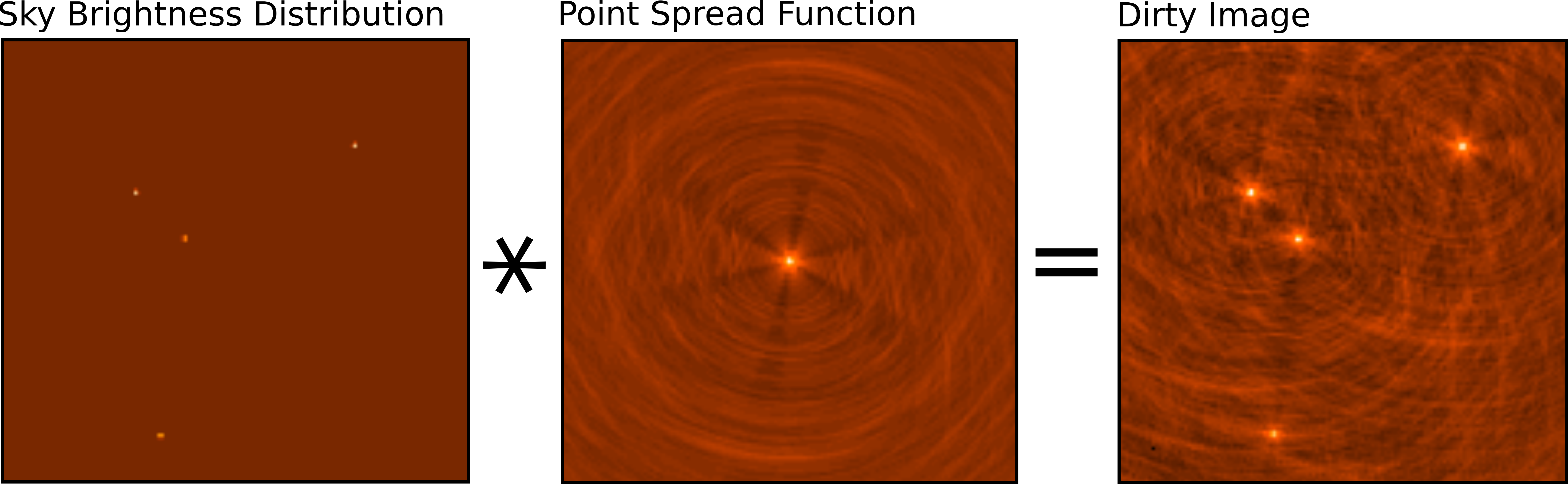}
\caption{Convolution of the sky brightness distribution (left) with the instrumental point-spread-function (middle) results in the `dirty' image (right). This example is based on simulated VLA data.}
\label{fig:convolution}
\end{figure*}

\subsubsection{Single-dish calibration}
\label{sec:sdcalibration}

The concept of single-dish calibration is different from that of an interferometer. Atmospheric variability that causes phase decoherence in interferometric data is not relevant for single-dish observations, but large-scale atmospheric fluctuations that are resolved out by interferometers yield power in single-dish telescopes. Single-dish telescopes detect and quantify signals in brightness temperature $T_{\rm B}$ (Kelvin). The brightness temperature of an astronomical target can be measured through:
\begin{equation}
\frac{T_{\rm target}}{T_{\rm sys}}\,=\,\frac{T_{\rm ON}\,-\,T_{\rm OFF}}{T_{\rm OFF}},
\end{equation}
where $T_{\rm ON}$ is the on-target measurement, and $T_{\rm OFF}$ is a measurement of the blank sky, taken at roughly the same elevation and time as the target measurement, but ideally absent of any target emission or contaminating sources at the frequencies of interest to the observer. The measured signal includes contributions from sky targets (for $T_{\rm ON}$), cosmic background signals, atmosphere, ground, and the telescope itself, as well as noise from the instrument's electronics. $T_{\rm sys}$ is the system temperature, which is obtained through an observation of the sky together with measurements of loads with known temperatures that are placed in front of the receiver. 

Typical observing modes include Position Switching, by which separate exposures are taken at discrete `ON' and `OFF' positions on the sky, or On-The-Fly mapping, where the telescope smoothly scans the field that contains the `ON' and `OFF' measurements \citep[e.g.,][]{one02,man07,saw08}. In CASA, the calibration of single-dish data generally requires several steps, namely the application of the $T_{\rm sys}$ calibration, and of the `sky' calibration, i.e., $T_{\rm OFF}$. The $T$ values are a function of frequency, hence calibration must be performed in the spectral domain. In addition, in the spectral domain, single-dish data rely on accurate fitting and subtraction of the spectral baseline emission.

Single-dish data in CASA rely on the same MeasurementSet as interferometric data. For single-dish calibration, various CASA tasks can be used that are also suitable for interferometry, such as \casatask{listobs}, \casatask{flagdata}, \casatask{gencal}, and \casatask{applycal}. In addition, a number of dedicated single-dish tasks are available in CASA, including \casatask{sdcal} for calibration, \casatask{sdgaincal} for removing time-dependent gain variations, and \casatask{sdbaseline} for fitting and subtracting a baseline from single-dish spectra.

\subsection{Data processing IV: Data manipulation}
\label{sec:manipulation}

CASA contains a number of tasks that allow manipulation of visibility data. This includes tasks to concatenate, average, split, weight, or regrid data in various ways (e.g., \casatask{concat}, \casatask{split}, \casatask{statwt}, and \casatask{cvel}), as well as dedicated tasks for Hanning smoothing (\casatask{hanningsmooth}), continuum subtraction (\casatask{uvcontsub}), model subtraction (\casatask{uvsub}), correcting the astronomical positions of targets (\casatask{fixplanets}), or shifting the interferometric phase center (\casatask{phaseshift}). While historically each step in the manipulation of  visibility data was done by individual tasks, currently the multi-purpose task \casatask{mstransform} combines most of the above mentioned functionality, with the possibility of applying each of these transformations separately or together in an in-memory pipeline, thus avoiding unnecessary input/output (I/O) steps. Most of the manipulation of visibility data is done after calibration, but before imaging.

\subsection{Data processing V: Imaging \& Deconvolution}
\label{sec:imaging}

\subsubsection{Interferometric imaging}
\label{sec:intimaging}

Image reconstruction in radio interferometry \citep{rea78,sra89,cor95,rau09,fre09} is the process of solving the linear system of equations:
\begin{equation}
\vec{V}^{\rm TRUE}\,=\,[A]\vec{I}
\label{eqn:image}
\end{equation}
where $\vec{V}^{\rm TRUE}$ represents visibilities that have been calibrated for direction independent effects through \casatask{applycal}, using the solutions from Eqn. \ref{eqn:measurementequation2}. $\vec{I}$ is a list of parameters that model the sky brightness distribution, such as an image consisting of pixels. [$A$] is a measurement operator that encodes the process of how visibilities are generated when a telescope observes a sky brightness $\vec{I}$, and is generally given by [$S_{dd}$][$F$], such that
\begin{equation}
\vec{V}^{\rm TRUE}\,=\,[S_{dd}][F]\vec{I}
\label{eqn:image}
\end{equation}
where [$F$] represents a 2D Fourier transform \citep{fou78,bra00} and [$S_{dd}$] represents a 2D spatial frequency sampling function that can include direction-dependent instrumental effects. An interferometer has a finite number of array elements, which means that [$A$] is not invertible because of unsampled regions of the ($u$,$v$)-plane. Therefore, this system of equations must be solved iteratively, applying constraints via various choices of image parameterizations and instrumental models. 
 
In CASA, this interferometric imaging process consists of converting a list of calibrated visiblities into a raw image, also called a `dirty' image, then `cleaning' this image to obtain an estimate of the true sky model by iterating through a few $\chi^2$ minimization steps that evaluate the goodness of fit of the current model with respect to the visibilities, and subsequently constructing the final image product. Under ideal conditions, the dirty image is the true sky brightness convolved with the point-spread-function (PSF) of the instrument and added noise terms (Fig. \ref{fig:convolution}). The PSF typically has a complex shape determined by the ($u$,$v$)-coverage and interference pattern of the --often sparsely populated-- distribution of antennas in the interferometer. Moreover, in reality, the dirty image is based only on the visibilities that are sampled by the interferometer, hence it does not contain the complete information about the true sky brightness distribution. There are several stages to forming a dirty image from interferometric data, including weighting, convolutional resampling, Fourier transformation, and normalization. To obtain the final `clean' image of the source, a model of the true sky brightness distribution is reconstructed from the dirty image and the PSF in a process called `deconvolution' \citep{cor89}, and subsequently convolved with a Gaussian that represents the instrumental resolution specified by the main lobe of the PSF. In this Section, we explain those concepts of imaging, weighting, gridding, and deconvolution that are most sensitive to parameters that need to be specified by users in CASA. This includes an overview of how deconvolution and image restoration is implemented in the CASA imaging task \casatask{tclean}.

\subsubsection{Weighting schemes}
\label{sec:weighting}

As part of the imaging process, the visibilities can be weighted in different ways to alter the instrument's natural response function, affecting resolution, sidelobe suppression, and root-mean-square (rms) noise levels. There are three main weighting schemes for radio interferometric data: natural, uniform, and Briggs (robust) weighting \citep{bri95}.

\begin{itemize}
\item{\casaparm{natural} {\sl weighting}: The data are gridded into ($u$,$v$)-cells for imaging, with weights given by the visibility weights in the MeasurementSet. This results in a higher imaging weight for higher ($u$,$v$)-density. Natural weighting produces an image with the lowest noise, but often with lower than optimal resolution. } 
\item{\casaparm{uniform} {\sl weighting}: The data are first gridded to a number of cells in the ($u$,$v$)-plane, and afterwards the ($u$,$v$)-cells are re-weighted to have uniform imaging weights. Uniform weighting produces an image with higher resolution, but which has increased noise compared to \casaparm{natural} or \casaparm{briggs} weighting. Also available in CASA is the option of \casaparm{superuniform} weighting, which increases the number of cells to define the ($u$,$v$)-plane patch for the weighting renormalization, and by doing so further increases the resolution and noise.} 
\item{\casaparm{briggs} (\casaparm{robust}) {\sl weighting}: This is a flexible weighting scheme that provides a trade-off between resolution and sensitivity \citep{bri95}. It uses a robustness parameter that takes values between -2.0 (close to \casaparm{uniform} weighting) and 2.0 (close to \casaparm{natural} weighting). CASA also includes two modified versions of the Briggs weighting scheme, named \casaparm{briggsabs} and \casaparm{briggsbwtaper}. }
\item{\casaparm{uvtaper:} Optionally, a multiplicative Gaussian taper can be applied to the spatial frequency grid, in addition to any of the above options. This effectively downweights the longer baselines, decreasing the resolution compared to the original weighting scheme it is applied on top of, and increasing surface-brightness sensitivity.}
\end{itemize}

\subsubsection{Imaging mode and Gridding}
\label{sec:gridding}

After visibilities are properly weighted, the next steps in the imaging process of interferometric data consists of gridding, Fourier transformation \citep{fou78,bra00}, and normalization of the data. Imaging weights and weighted visibilities are first resampled using gridding convolution functions onto a regular ($u$,$v$)-grid in a process called convolutional resampling \citep{bro75}. The result is Fourier-inverted using a fast-Fourier transformation, and subsequently grid-corrected to remove the image-domain effect of the gridding convolution function. 

In CASA, the type and shape of the image product is determined by various parameters, including data selection, definition of the spectral mode, and the \casaparm{gridder} algorithm. Image products can be continuum images, spectral cubes, or Stokes images, while multiple pointings can be captured into a mosaic image. The standard gridding in CASA relies on prolate spheroidal functions, along with image-domain operators to correct for direction-dependent effects. More sophisticated gridder modes can apply direction-dependent, time-variable and baseline-dependent corrections during gridding in the visibility-domain, by computing the appropriate gridding convolution kernel to use along with the imaging-weights. These gridder modes are superior for certain data sets, like widefield and wideband data, but are more computationally intensive.

Typical imaging modes in CASA include:
\begin{itemize}
\item{\casaparm{mfs}{\sl, multi-frequency synthesis imaging:} Standard continuum imaging, where selected data channels across a (wide) range in frequencies are mapped onto a single wideband image \citep[][]{con90}.}
\item{\casaparm{cube} {\sl imaging:} Standard imaging of spectral lines, where selected data channels are mapped to a number of image channels using various interpolation schemes. Optionally, the image cube can be corrected for Doppler effects and ephemeris tracking.}
\item{\casaparm{stokes} {\sl imaging:} Imaging of different types of polarization products, often related to Stokes {\sl I}, {\sl Q}, {\sl U}, {\sl V}, or combinations thereof \citep{sto52,ham96_stokes}.}
\item{\casaparm{mosaic} {\sl imaging:} Continuum or spectral-line imaging of observations that consist of multiple pointings, producing fields-of-view larger than the primary beam. Options for stitched and joint mosaics are available \citep[e.g.,][]{eke79,cor88,bre94,con94,sau96_2}.}
\item{\casaparm{mtmfs}{\sl, multi-term multi-frequency synthesis imaging:} Wideband image reconstruction that produces Taylor-coefficient maps, which represent smoothly varying spectral structure across the field-of-view of the image (\citealt{rau11}, see also \citealt{sau94} \citealt{lik05}).}
\item{{\sl imaging multiple (outlier) fields:} Large fields imaged as a main field plus multiple (smaller) outlier fields, rather than visibility data being gridded onto one large ($u$,$v$)-grid.}
\item{{\sl correction for widefield and wideband instrumental effects:} High dynamic range imaging of data with a large field-of-view and large fractional bandwidth, by accounting for a number of widefield and wideband effects during gridding. These effects include sky curvature, non-coplanar baselines \citep{cor92}, and antenna-based aperture illumination functions that change with time, frequency, and polarization \citep{sek19}.} 
\end{itemize}

A-Projection is an example of an advanced gridding algorithm in CASA, which can correct for complex widefield \citep{bha08} and wideband \citep{bha13} effects \citep[see also][]{tas13}. This can be critical for sensitive widefield continuum observations, for example those performed using the lower frequency bands of the VLA \citep[e.g.,][]{rau16,sch19}. The aperture illumination function results in a direction-dependent complex gain that causes the primary beam to vary with time, frequency, polarization, and antenna \citep{jag18a, sek19}. These variations may be caused by pointing errors, feed leg structures that break azimuthal symmetry, parallactic angle rotation, and varying dish sizes. These variations in the primary beam are corrected during the gridding using the wideband A-Projection algorithm, which computes gridding convolution functions for each baseline as the convolution of the complex conjugates of two antenna aperture illumination functions \citep{bha13}. In addition, sky curvature and non-coplanar baselines result in a so-called w-term that is not zero, which means that standard 2D imaging applied to such data will produce artifacts around sources away from the phase center. A W-Projection algorithm corrects for the effects introduced by the w-term, using gridding convolution functions that are computed based on the Fourier transform of the Fresnel electro-magnetic wave propagator for a finite set of w-values \citep{cor08_wterm}. The \casaparm{wproject}, \casaparm{mosaic}, and \casaparm{awproject} gridding options implement different approximations and combinations of the A- and W-Projection algorithms.

\subsubsection{Deconvolution and image reconstruction}

Deconvolution refers to the process of reconstructing a model of the sky brightness distribution, given a dirty image and the PSF of the instrument. This process is called a deconvolution, because under ideal conditions, the dirty image can be written as the result of a convolution of the true sky brightness and the PSF of the instrument (Fig. \ref{fig:convolution}). The concept of deconvolution is a widely used technique in signal and image processing, and explaining the fundamentals is beyond the scope of the current paper \citep[see][]{cor89}. Instead, we will give an overview of how the technique of deconvolution and image reconstruction is practically implemented in CASA through the task \casatask{tclean}.

\paragraph{Tclean, CASA's powerful imaging task}

\begin{figure*}[htb]
\centering
\includegraphics[width=0.8\textwidth]{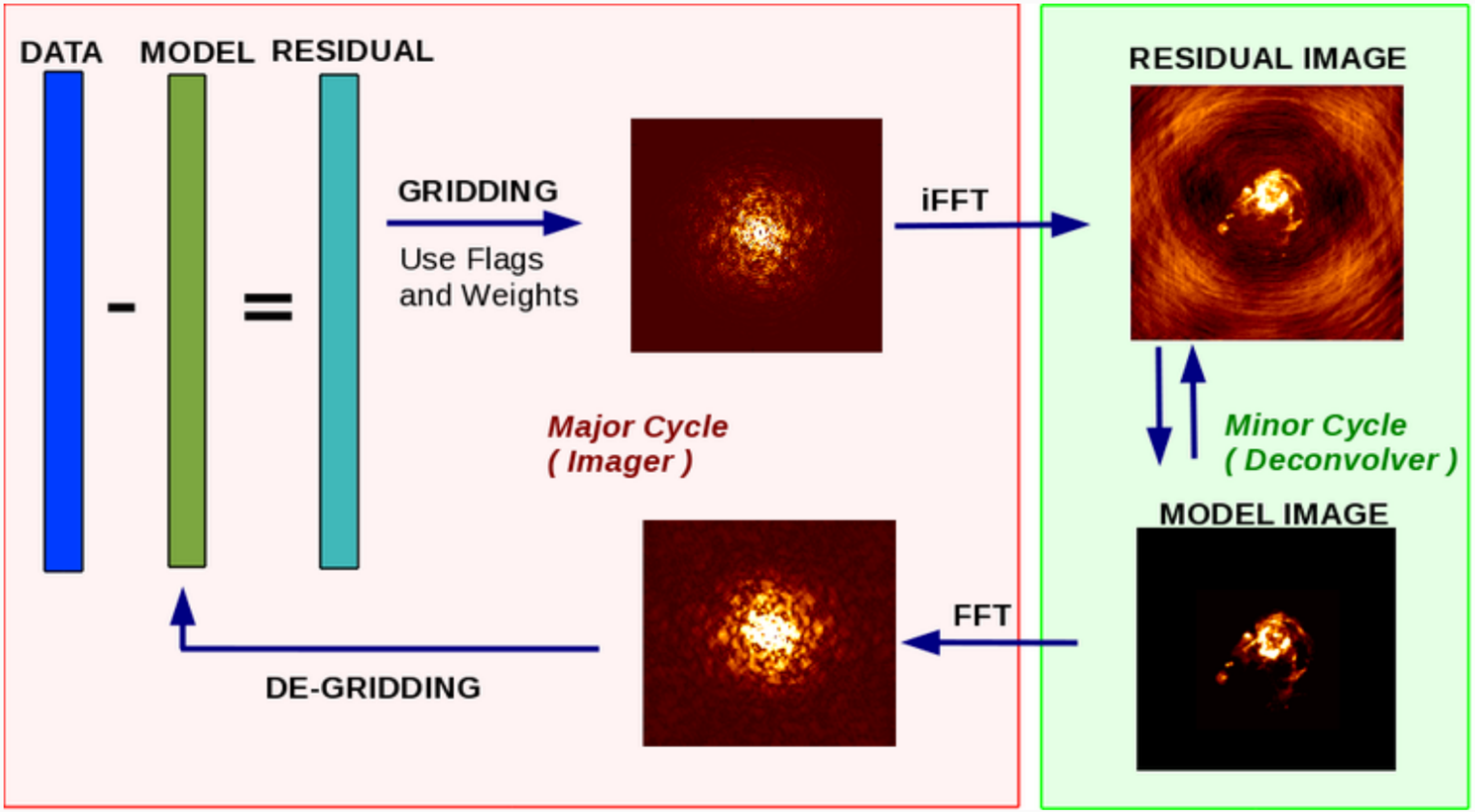}
\caption{Schematic overview of the iterative image reconstruction in the CASA task \casatask{tclean}. The left (red) and right (green) boxes show the major and minor cycle, respectively. After gridding the visibilities and applying an inverse fast Fourier transform (iFFT), deconvolution on the (residual) image is done in the minor cycle. At the end of the minor cycle, the model image is translated back into the ($u$,$v$)-domain with a fast Fourier transform (FFT), where it is subtracted from the visibility data at the start of the next major cycle.}
\label{fig:tclean}
\end{figure*}

The CASA task used for imaging is \casatask{tclean}, which is based on the CLEAN algorithm \citep{hog74,sch78}. The \casatask{tclean} task takes the calibrated visibilities from the MS and applies weighting, sampling, Fourier transformation, deconvolution, and image restoration according to inputs specified by the user or pipeline. The output of \casatask{tclean} is a reconstructed image of the astronomical source. In addition, \casatask{tclean} by default also produces various other images that may be useful for subsequent processing or analysis, including images of the derived sky model and residuals, instrumental PSF, primary beam response, and sum of the data weights.

Image reconstruction in CASA follows the CLEAN method by Cotton $\&$ Schwab \citep[see][]{sch84}, which consists of an outer loop of major cycles and an inner loop of minor cycles (Fig.\,\ref{fig:tclean}). The major cycle implements transforms between the visibility data and image domain, while the minor cycle represents the deconvolution step that operates purely in the image domain. This method implements an iterative weighted $\chi^{2}$ minimization that solves the measurement equation, but allows for a practical trade-off between the efficiency of operating in the image domain and the accuracy that comes from frequently returning to the ungridded list of visibilities. It also allows for minor cycle algorithms to have their own internal optimization scheme.

The minor cycle performs the deconvolution step by separating sky emission from the PSF and building up a model of the true sky brightness distribution. Different algorithms are available in CASA to identify flux components that are to be included in the sky model. These are \casaparm{hogbom}, \casaparm{clark}, and \casaparm{clarkstokes}, which all identify model-components as delta-functions \citep{hog74,cla80}; \casaparm{mem}, which is based on the Maximum Entropy Method \citep{cor85}; \casaparm{multiscale}, which assigns model components at different spatial scales and is useful for images that include extended emission \citep{cor08}; \casaparm{asp}, which is an Adaptive-Scale Pixel algorithm for more flexible multi-scale source modeling \citep{bha04}; and \casaparm{mtmfs} or Multi-Term (Multi-Scale) Multi-Frequency Synthesis, which is a multi-scale and multi-term cleaning algorithm optimized for wideband imaging \citep{rau11}. Deconvolution begins with a residual image that includes the astronomical signal, PSF, and noise. In this residual image, model components are identified using one of the above algorithms. After the algorithm identifies a model component, the model is updated accordingly and the effect of the PSF is removed by subtracting a scaled PSF from the image at the location of each component. Many such iterations of finding peaks and subtracting PSFs form the minor cycle (Fig.\ref{fig:tclean}). This deconvolution step can be performed for the whole image, or across regions of the image as defined with a mask. At the start of the next major cycle, the model of the sky brightness that is derived during the minor cycle is evaluated against the measurement equation and converted into a predicted list of model visibilities, which are then subtracted from the data to form a new residual image.

During the major cycle, a pseudo inverse of [$S_{dd}$][$F$] is computed and applied to the visibilities. Operationally, weighted visibilities are convolutionally resampled onto a grid of spatial-frequency cells, inverse Fourier transformed, and normalized via user-specified inputs. The accuracy and efficiency of the image reconstruction depends on the algorithms chosen for [$S_{dd}$] and [$F$], and direction-dependent instrumental effects can be accounted for via carefully constructed convolution functions in CASA (see Sect.\,\ref{sec:gridding}). 

Image deconvolution, and the corresponding sequence of major and minor cycles, continues until a user-specified threshold criteria has been reached across the selected region. This threshold can be specified based on a cutoff level or number of interactions, optionally combined with a mask. Such a mask can be constructed a-priori by the user, iteratively after each major cycle, or determined automatically using the \casaparm{auto-multithresh} algorithm. The \casaparm{auto-multithresh} option evaluates the noise and sidelobe thresholds in the residual image to set an initial mask at the start of each minor cycle and then cascades that mask down to lower signal-to-noise, taking into account the fundamental properties of the image \citep{kep20}.

After deconvolution, the output sky model is restored by a Gaussian function that represents the instrumental resolution specified by the PSF main lobe, but without the sidelobes. This results in a cleaned image of the sky. 

The final image can optionally be corrected for the response of the primary beam of the telescope, in case this has not already been done during the gridding stage. This will provide true values of the flux density throughout the field-of-view. Options for wideband primary-beam correction are included within the \casaparm{mosaic} and \casaparm{awproject} gridders.

\subsubsection{Single-dish imaging}

Converting single-dish observations into an image or cube is done almost entirely in the image domain. The single-dish data should be calibrated, according to the process described in Sect.\,\ref{sec:sdcalibration}. The CASA task \casatask{tsdimaging} then converts the single-dish observations into an image or cube by forming and populating the image grid. The convolution kernels that can be used for gridding the image in CASA consist of a boxcar function, Gaussian function, primary beam function, prolate spheroidal wave function, and Gaussian-tapered Bessel function \citep[][]{man07}. 
The \casatask{tsdimaging} task by default chooses reasonable values for parameters like pixel size, image dimensions, and spectral resolution, but all these can be adjusted manually in CASA.

\subsubsection{Image combination}

Combining interferometric with single-dish data allows for reconstructing the flux of the astronomical source on all spatial scales, including extended emission that is resolved out even by the shortest baselines of the interferometer. Techniques that rely on joint deconvolution \citep{cor88,sau96_2}, either in the image domain \citep{sta99,sta02,pet10,jun16,rau19} or in the visibility domain \citep{kur09,kod11,kod19,teu19}, have been shown to be fruitful for combining single-dish and interferometric data. 

CASA contains a dedicated joint deconvolution algorithm for wideband multi-term and mosaic imaging, captured in the CASA task \casatask{sdintimaging} \citep{rau19}. Interferometer data are gridded into an image cube and corresponding PSF. The single-dish image and PSF cubes are combined with the interferometer cubes in a feathering step \citep{cot17}. This feathering step is based on the CASA task \casatask{feather}, which Fourier transforms the single-dish and interferometric images to a gridded visibility plane and weights them by their spatial frequency response. The joint image and PSF cubes then form inputs to any deconvolution algorithm, be it multi-channel (\casaparm{cube}), multi-frequency synthesis (\casaparm{mfs}), or multi-term mfs (\casaparm{mtmfs}) mode. Model images from the deconvolution algorithm are translated back to model image cubes prior to subtraction from both the single-dish image cube and the interferometer data to form a new pair of residual image cubes to be feathered in the next iteration. In the case of wideband mosaic imaging, wideband primary beam corrections are always performed per channel of the image cube, followed by a multiplication by a common primary beam, prior to deconvolution.

The \casatask{sdintimaging} task supports joint deconvolution for spectral cubes as well as multi-term wideband imaging, operates on single pointings as well as joint mosaics, includes corrections for frequency dependent primary beams, and optionally allows the deconvolution of only single-dish images in both \casaparm{cube} and (multi-term) \casaparm{mfs} mode. An option has also been provided to tune the relative weighting of the single-dish and interferometer data alongside the standard weighting schemes used for interferometric imaging.

\subsection{Data processing VI: Analysis \& Visualization}
\label{sec:analysis}

CASA also offers a suite of features, mostly in the form of Graphical User Interfaces (GUIs), for the visualization and analysis of radio astronomical data. Various GUIs are available to inspect the raw data and meta-data from the telescopes, including \casatask{browsetable} to display CASA tables, \casatask{plotants} to plot the antenna positions, and \casatask{plotms} to visualize the visibility data (see Sect.\,\ref{sec:flagging}).

An new tool for visualizing image products is the {\sl Cube Analysis and Rendering Tool for Astronomy} (CARTA; \citealt{com21}),\footnote{\url{https://cartavis.org}} which is an external image visualization and analysis software designed for ALMA, the VLA, and pathfinder telescopes for the Square Kilometre Array (SKA; \citealt{laz09}). CARTA is developed by a consortium of the Academia Sinica Institute of Astronomy and Astrophysics (ASIAA), the South African Inter-University Institute for Data Intensive Astronomy (IDIA), the National Radio Astronomy Observatory (NRAO), and the Department of Physics at the University of Alberta. CARTA is being developed bearing in mind the increasing demands of next-generation radio telescopes, and will be a standard visualization tool that can be used for data processed with CASA.

\subsection{Data processing VII: Simulations}
\label{sec:simulations}

The capability of simulating observations and data sets from the VLA, ALMA, and other existing and future observatories is an important use-case for CASA. This not only gives users a better understanding of the scientific capabilities and expected output of these telescopes, but also provides benchmarks to test the performance, optimization, and reproducability of the CASA software. CASA can create simulated MeasurementSets for any interferometric array. For a large number of interferometers, array configuration files of the antenna distributions are readily available in CASA. The tasks available for simulating observations in CASA are:
\begin{itemize}
\item{\casatask{simobserve:} create simulated MeasurementSets for an interferometric or total power observation with a specific telescope.}
\item{\casatask{simanalyze:} image and analyze simulated MeasurementSet data, including diagnostic images and plots.}
\item{\casatask{simalma:} Streamlined combination of \casatask{simobserve} and \casatask{simanalyze} for ALMA data. The \casatask{simalma} task can simulate ALMA observations in one go, including multiple configurations of the main 12-m interferometric array, the 7-m Atacama Compact Array (ACA; \citealt{igu09}), and total power measurements.}
\end{itemize}
In addition to these simulation tasks, the \casatask{simulator} methods as part of the CASA tools allow for even greater flexibility and functionality, especially for non-ALMA use-cases. For example, the simulator tools can be used to calculate and apply calibration tables that represent some of the most important corrupting factors, such as atmospheric and instrumental effects.

\subsection{Very Long Baseline Interferometry}
\label{sec:vlbi}

CASA can also process Very Long Baseline Interferometry (VLBI) data \citep[e.g.,][]{wie11}, including data from the Very Long Baseline Array (VLBA; \citealt{kel85}) and European VLBI Network (EVN; \citealt{zen15}). The CASA package contains the VLBI-specific tasks \casatask{fringefit} and \casatask{accor} \citep{bem19}. The \casatask{fringefit} task determines phase, delay, delay-rate, and optionally dispersive delay solutions, as a function of time and spectral window. This enables correcting the visibility phases for errors introduced by the atmosphere, the signal paths of the instrument, or other pre-calibration factors. The \casatask{accor} task determines the amplitude corrections from the apparent normalization of the mean autocorrelation spectra. This corrects for errors in sampler thresholds during an observation, as caused by the DiFX correlator used by the VLBA \citep{nap94}. In addition to \casatask{fringefit} and \casatask{accor}, other tasks that are used for processing normal interferometric data can also handle VLBI data. 

Externally, \citet{jan19} built the first generic VLBI calibration and imaging pipeline on top of CASA. This $\rpicard$ VLBI pipeline formed a critical component in the processing of data from the Event Horizon Telescope, which resulted in the first image of a black hole \citep[e.g.,][]{eve19a,eve19b,jan19}

VLBI development for CASA is being led by JIVE in collaboration with NRAO, and is an ongoing effort. The VLBI-specific tools, tasks, and their operation are described in detail in a CASA-VLBI paper by \citet{bem22}, which complements this current CASA paper.

\subsection{Algorithm Research $\&$ Development}
\label{sec:RandD}

The features and algorithms within CASA are continually improved as use-cases and needs evolve. Algorithm Research and Development (R$\&$D) is performed in conjunction with the NRAO Algorithm R$\&$D Group, with scientific staff engaged in telescope and pipeline operations, and within the CASA development team. Algorithms that are expected to be incorporated into CASA in the next few years include a wideband version of the Adaptive Scale Pixel (ASP) deconvolution algorithm \citep{bha04,hsi21}, Full Mueller imaging for wide bandwidths \citep{jag18b}, an automated flagging algorithm based on binning of the ($u,v$)-data, and a Graphics Processing Unit (GPU) implementation of the widefield A-Projection algorithm \citep{pok21}. 

The CASA team is also developing a new set of GUIs for examining, processing, and simulating data in CASA. Such new GUIs will allow for better integration in Python, and offer usability within Notebook environments \citep{klu16}.

\section{ALMA and VLA pipelines}
\label{sec:pipelines}

One of the core aspects of CASA development is to support various pipelines for the operation of ALMA and the VLA. ALMA maintains both a calibration and an imaging pipeline, which are built on top of CASA (Hunter et al. in prep.; see also \citealt{mud14}; \citealt{gee19}; \citealt{mas20}). These ALMA pipelines are used for automated and customized processing of ALMA data. Different versions of these pipelines are maintained by NRAO for processing data from the VLA and VLA Sky Survey \citep{ken20}, as well as the Science Ready Data Products program \citep{lac20}. The pipelines use standard CASA tasks and tool methods, as well as custom-made pipeline tasks and analysis utilities. The ALMA and VLA pipelines, and associated pipeline tasks, are bundled only with select versions of CASA.

CASA also supports the pipeline for processing single-dish data from the Nobeyama 45-m telescope obtained in On-The-Fly observing mode. In addition, the flexibility of the CASA software also allows for creation of custom build tasks and external pipelines, such as the VLBI pipelines mentioned in Sect. \ref{sec:vlbi}.

\section{Development Process}
\label{sec:devprocess}
CASA's development planning is done in conjunction with a stakeholder group containing representatives from current NRAO and ALMA telescopes, pipeline-enabled projects and external users. VLBI and Single-Dish stakeholder inputs are coordinated via our partner developer teams and included in our overall development plans. Stakeholder priorities are then balanced with needs for software maintenance and evolution to prioritize work on a half-yearly timescale, while CASA puts out regular releases on a few-month cadence.

For each feature that is added, standard software development practices are employed, starting with the definition of requirements and specifications prior to development, and ending with internal verification and external validation stages. Verification tests include functional and unit tests written against specifications and aimed at code and interface coverage. We also maintain stakeholder-verification tests targeted towards specific dominant usage modes (i.e. operational pipelines) and are developing performance and benchmarks test-suites. Validation is typically done by external science staff and evaluates applicability to stakeholder use-cases. Some numerical or algorithmic features undergo additional detailed characterization, either by the development team or by external testers, to assess the effectiveness of the implementation for the intended use-cases.

\section{Next-Generation CASA}
\label{sec:ngcasa}

With the massive increase in data rates and volumes that are expected in the era of the next-generation radio telescopes, such as the next-generation VLA (ngVLA; \citealt{mur18}) and ALMA wideband sensitivity upgrade \citep{bro19}, a next-generation of data processing software needs to be efficient and easily scalable to large computing environments. A design phase has started for a next-generation CASA software, built on top of CASA's next-generation infrastructure. This next-generation CASA is aimed at reducing code complexity and development time, while at the same time increasing reliability, flexibility, and scalability. A prototype implementation of the CASA next-generation infrastructure, currently referred to as CNGI,\footnote{\url{https://cngi-prototype.readthedocs.io}} uses the Zarr storage system \citep{mil20}, the Xarray API \citep{hoy17} and the Dask parallel processing framework \citep{das16}.

\section{Conclusions}
\label{sec:conclusions}

The Common Astronomy Software Applications (CASA) is a versatile software package for the calibration, imaging, and analysis of data from ALMA, the VLA, and other radio telescopes. CASA aims at maintaining full functionality for the processing of astronomical data produced by aperture-synthesis arrays and single-dish radio telescopes. With comprehensive documentation and a stakeholder-based development process, CASA serves the astronomical community as a leading software for the processing of radio data. With its core aspect of supporting pipelines, and 
its initiative of developing a modernized software infrastructure to address increasing data rates and scalability, CASA will continue to meet the challenges of handling the ever-increasing complexity of data produced by current and next-generation radio telescopes.

\begin{acknowledgments}
The CASA team is indebted to all our colleagues who have previously been associated with and contributed to the CASA project, and we thank them for their dedicated work over the years. CASA is being developed by an international consortium of scientists and software engineers based at the National Radio Astronomical Observatory (NRAO), the European Southern Observatory (ESO), the National Astronomical Observatory of Japan (NAOJ), and the Joint Institute for VLBI European Research Infrastructure Consortium (JIV-ERIC), under the guidance of NRAO. The CASA team acknowledges its international partners the Academia Sinica Institute of Astronomy and Astrophysics (ASIAA), the CSIRO division for Astronomy and Space Science (CASS), the Netherlands Institute for Radio Astronomy (ASTRON), the Inter-University Institute for Data Intensive Astronomy (IDIA), and the University of Alberta for 3rd-party software (CARTA/Casacore) that CASA both uses and contributes too. We also thank our colleagues at the ALMA, VLA, VLASS, SRDP, and Nobeyama pipeline development teams and science working groups, the ALMA regional centers, and the VLA, VLBA, and Nobeyama instrument teams for their valuable help with code validation, quality control, and user support. The CASA Team is indebted to NRAO’s Scientific Computing Group lead by James Robnett, and thankful for IT support by the Computing $\&$ Information Services group. We are also grateful for the feedback from the CASA Users Committee and other stakeholders in support of the CASA project. Finally, we sincerely thank the anonymous referee for valuable suggestions that improved this paper. The National Radio Astronomy Observatory is a facility of the National Science Foundation operated under cooperative agreement by Associated Universities, Inc. ALMA is a partnership of ESO (representing its member states), NSF (USA) and NINS (Japan), together with NRC (Canada), MOST and ASIAA (Taiwan), and KASI (Republic of Korea), in cooperation with the Republic of Chile. The Joint ALMA Observatory is operated by ESO, AUI/NRAO and NAOJ. The European VLBI Network is a joint facility of independent European, African, Asian, and North American radio astronomy institutes. This work is supported by the ERC Synergy Grant “BlackHoleCam: Imaging the Event Horizon of Black Holes” (Grant 610058). SS acknowledges financial support from the Inter-University Institute for Data Intensive Astronomy (IDIA). IDIA is a partnership of the University of Cape Town, the University of Pretoria, the University of the Western Cape and the South African Radio astronomy Observatory.
\end{acknowledgments}

%



\software{CASA (\url{https://casadocs.readthedocs.io}), Casacore \citep{cas19}, CARTA \citep{com21}, AIPS \citep{gre03}, AIPS++ \citep{mcm06}, Python \citep{ros95}, IPython \citep{per07}, C++ \citep{str97}, Fortran \citep{bac56}.}

\end{document}